\documentclass[letterpaper, 10 pt, conference]{ieeeconf}

\newtheorem{theorem}{Theorem}[section]
\newtheorem{cor}[theorem]{Corollary}

\newtheorem{prop}[theorem]{Proposition}

\newtheorem{problem}{Problem}
\newtheorem{definition}[theorem]{Definition}
\newtheorem{rem}[theorem]{Remark}

\newcommand{\set}[1]{\left\{#1\right\}}

\IEEEoverridecommandlockouts
\usepackage{ragged2e}
\usepackage{cancel}
\usepackage{microtype}
\usepackage[T1]{fontenc}
\usepackage[utf8]{inputenc}
\usepackage{tabularx,ragged2e,booktabs,caption}
\newcolumntype{C}[1]{>{\Centering}m{#1}}

\usepackage{graphicx} 
\graphicspath{{figures/}}
\usepackage{hyperref}
\hypersetup{
	colorlinks=true,
	linkcolor=black,
	filecolor=black,      
	urlcolor=black,
	citecolor=black,}
\usepackage{epsfig} 
\usepackage{ragged2e}
\usepackage{amsmath}
\usepackage{amssymb}
\usepackage{epstopdf}
\usepackage{cite}
\usepackage[algo2e,linesnumbered,ruled,lined]{algorithm2e}
\SetKwComment{Comment}{$\triangleright$\ } {}
\usepackage{multirow}
\usepackage{rotating}
\usepackage{subfigure}
\usepackage{color}
\usepackage{mysymbol}
\usepackage{url}
\usepackage{tikz}
\usepackage{flushend}
\usepackage{microtype} 
\usetikzlibrary{fit,automata,positioning,angles,quotes}
\usepackage{float}
\restylefloat{figure}
\usepackage{dsfont}
\usepackage{inconsolata}
\newcommand{\hosein}{\textcolor{black}}

\begin{document}
	\title{\LARGE \bf  Reinforcement Learning for Temporal Logic Control Synthesis\\ with Probabilistic Satisfaction Guarantees}
	
	\author{M. Hasanbeig, Y. Kantaros, A. Abate, D. Kroening, G. J. Pappas, I. Lee \thanks{M. Hasanbeig, A. Abate, and D. Kroening are with the Department of Computer Science, University of Oxford, UK, and are supported by the ERC project 280053 (CPROVER) and the H2020 FET OPEN 712689 SC$^2$. $\left\{\text{hosein.hasanbeig,aabate,kroening}\right\}$@cs.ox.ac.uk. Y. Kantaros, G. J. Pappas, and I. Lee are with the School of Engineering and Applied Science (SEAS), University of Pennsylvania, PA, USA, and are supported by the AFRL and DARPA under Contract No. FA8750-18-C-0090 and the ARL RCTA under Contract No. W911NF-10-2-0016 $\left\{\text{kantaros,pappasg,lee}\right\}$@seas.upenn.edu.}}
	
	\maketitle
	
	\begin{abstract}
		Reinforcement Learning (RL) has emerged as an efficient method of choice for solving complex sequential decision making problems in automatic control, computer science, economics, and biology. In this paper we present a model-free RL algorithm to synthesize control policies that maximize the probability of satisfying high-level control objectives given as Linear Temporal Logic (LTL) formulas. Uncertainty is considered in the workspace properties, the structure of the workspace, and the agent actions, giving rise to a Probabilistically-Labeled Markov Decision Process (PL-MDP) with unknown graph structure and stochastic behaviour, which is even more general case than a fully unknown MDP. We~first translate the LTL specification into a Limit Deterministic B\"uchi Automaton (LDBA), which is then used in an on-the-fly product with the PL-MDP. Thereafter, we define a synchronous reward function based on the acceptance condition of the LDBA. Finally, we show that the RL algorithm delivers a policy that maximizes the satisfaction probability asymptotically. We~provide experimental results that showcase the efficiency of the proposed method. 
		%
	\end{abstract}
	
	\section{Introduction} 
	The use of temporal logic has been promoted as formal task specifications for control synthesis in Markov Decision Processes (MDPs) due to their expressive power, as they can handle a richer class of tasks than the classical point-to-point navigation. Such rich specifications include safety and liveness requirements, sequential tasks, coverage, and temporal ordering of different objectives
	\cite{fainekos2005hybrid,kress2009temporal,bhatia2010sampling,kantaros2017Csampling,kantaros2017distributed}. 
	%
	%
	Control synthesis for MDPs under Linear Temporal Logic (LTL) specifications has also been studied in \cite{ding2011mdp,wolff2012robust,ding2014optimal,guo2018probabilistic,tkachev}. 
	Common in these works is that, in order to synthesize policies that maximize the satisfaction probability, exact knowledge of the MDP 
	is required.  Specifically, these methods construct a product MDP by composing the MDP that captures the underlying dynamics with a Deterministic Rabin Automaton (DRA) that represents the LTL specification. Then, given the product MDP, probabilistic model checking techniques are employed to design optimal control policies \cite{baier2008principles,clarke1999model}. 
	
	In this paper, we address the problem of designing optimal control policies for MDPs with \textit{unknown} stochastic behaviour so that the generated traces satisfy a given LTL specification with maximum probability. Unlike previous work, uncertainty is considered both in the environment properties and in the agent actions, provoking a \textit{Probabilistically-Labeled} MDP (PL-MDP). This model further extend MDPs to provide a way to consider dynamic and uncertain environments. In order to solve this problem, we first convert the LTL formula into a Limit Deterministic B\"uchi Automaton (LDBA) \cite{sickert}. 
	It is known that this construction results in an exponential-sized automaton for LTL${}_{\setminus \text{GU}}$, and it results in nearly the same size as a DRA for the rest of LTL. LTL${}_{\setminus \text{GU}}$ is a fragment of linear temporal logic with the restriction that no until operator occurs in the scope of an always operator.
	On the other hand, the DRA that are typically employed in relevant work are doubly exponential in the size of the original LTL formula \cite{dra4}. Furthermore, a B\"uchi automaton is semantically simpler than a Rabin automaton in terms of its acceptance conditions \cite{sickert2,tkachev}, which makes our algorithm much easier to implement. 
	Once the LDBA is generated from the given LTL property, we construct on-the-fly a product between the PL-MDP and the resulting LDBA and then define a \textit{synchronous reward} function based on the acceptance condition of the B\"uchi automaton over the state-action pairs of the product. Using this algorithmic reward shaping procedure, a model-free RL algorithm is introduced, which is able to generate a policy that returns the maximum expected reward. Finally, we show that maximizing the expected accumulated reward entails the maximization of the satisfaction probability.
	
	\textbf{Related work} -- A model-based RL algorithm to design policies that maximize the satisfaction probability is proposed in \cite{topku,brazdil}. Specifically, \cite{topku} assumes that the given MDP model has unknown transition probabilities and builds a Probably Approximately Correct MDP (PAC MDP), which is composed with the DRA that expresses the LTL property. 
	The overall goal is to calculate the finite-horizon ($T$-step) value function for each state, such that the obtained value is within an error bound from the probability of satisfying the given LTL property. The PAC MDP is generated via an RL-like algorithm, then value iteration is applied to update state values. A similar model-based solution is proposed in \cite{dorsa}: this also hinges on approximating the transition probabilities, which limits the precision of the policy generation process. Unlike the problem that is considered in this paper, the work in~\cite{dorsa} is limited to policies whose traces satisfy the property with probability one. Moreover, \cite{topku,brazdil,dorsa} require to learn all transition probabilities of the MDP. As a result, they need a significant amount of memory to store the learned model~\cite{sutton}. This specific issue is addressed in \cite{wang2015temporal}, which proposes an actor-critic method for LTL specification that requires the graph structure of the MDP, but not all transition probabilities. The structure of the MDP allows for the computation of Accepting Maximum End Components (AMECs) in the product MDP, while transition probabilities are generated only when needed by a simulator. By contrast, the proposed method does not require knowledge of the structure of the MDP and does not rely on computing AMECs of a product MDP. A model-free and AMEC-free RL algorithm for LTL planning is also proposed in \cite{gao2019reduced}. Nevertheless, unlike our proposed method, all these cognate contributions rely on the LTL-to-DRA conversion, and uncertainty is considered only in the agent actions, but not in the workspace properties.
	
	In \cite{fulton} and \cite{fulton2} safety-critical settings in RL are addressed in which the agent has to deal with a heterogeneous set of MDPs in the context of cyber-physical systems. \cite{fulton3} further employs DDL \cite{ddl}, a first-order multi-modal logic for specifying and proving properties of hybrid programs. 
	
	The first use of LDBA for LTL-constrained policy synthesis in a model-free RL setup appears in \cite{hasanbeig2018logically,hasanbeig2018logically2}. Specifically, \cite{hasanbeig2018logically2} propose a hybrid neural network architecture combined with LDBAs to handle MDPs with continuous state spaces. The work in~\cite{hasanbeig2018logically} has been taken up more recently by~\cite{hahn}, which has focused on model-free aspects of the algorithm and has employed a different LDBA structure and reward, which introduce extra states in the product MDP. The authors also do not discuss the complexity of the automaton construction with respect to the size of the formula, but given the fact that resulting automaton is not a generalised B\"uchi, it can be expected that the density of automaton acceptance condition is quite low, which might result in a state-space explosion, particularly if the LTL formula is complex. As we show in the proof for the counter example in the Appendix-E the authors indeed have overlooked that our algorithm is episodic, and allows the discount factor to be equal to one. Unlike  \cite{hasanbeig2018logically,hasanbeig2018logically2,hahn}, in this work we consider uncertainty in the workspace properties by employing PL-MDPs.
	
	\textbf{Summary of contributions} -- \textit{First}, we propose a model-free RL algorithm to synthesize control policies for \textit{unknown} PL-MDPs which maximizes the probability of satisfying LTL specifications. \textit{Second}, we define a \textit{synchronous reward function} and we show that maximizing the accumulated reward maximizes the satisfaction probability.  \textit{Third}, we convert the LTL specification into an LDBA which, as a result, shrinks the state-space that needs to explored compared to relevant LTL-to-DRA-based works in finite-state MDPs. Moreover, unlike previous works, our proposed method does not require computation of AMECs of a product MDP, which avoids the quadratic time complexity of such a computation in the size of the product MDP~\cite{baier2008principles,clarke1999model}.  
	
	\section{Problem Formulation}\label{sec:problem}
	
	Consider a robot that resides in a partitioned environment with a finite number of states. To capture uncertainty in both the robot motion and the workspace properties, we model the interaction of the robot with the environment as a \textit{PL-MDP}, which is defined as follows.
	
	\begin{definition}[Probabilistically-Labeled MDP \cite{guo2018probabilistic}]\label{def:labelMDP}
		A~PL-MDP is a tuple $\mathfrak{M} = (\ccalX, x_0,  \ccalA, P_C, \mathcal{AP}, P_L)$, where $\ccalX$ is a finite set of states; $x_0\in\ccalX$ is the initial state; $\ccalA$ is a finite set of actions. With slight abuse of notation $\ccalA(x)$ denotes the available actions at state $x\in\ccalX$; 
		$P_C:\ccalX\times\ccalA\times\ccalX\rightarrow[0,1]$ is the transition probability function so that $P_C(x,a,x')$ is the transition probability from state $x\in\ccalX$ to state $x'\in\ccalX$ via control
		action $a\in\ccalA$ and $\sum_{x'\in\ccalS}P_C(x,a,x')=1$, for all $a\in\ccalA(x)$; $\mathcal{AP}$ is a set of atomic propositions; 
		and $P_L : \ccalX \times 2^{\mathcal{AP}}\rightarrow [0, 1]$ specifies the associated probability. 
		Specifically, $P_L(x,\ell)$ denotes the probability that $\ell\in 2^\mathcal{AP}$ is observed at state $x\in\ccalX$, where $\sum_{\ell\in2^{\mathcal{AP}}} P_L(x,\ell)=1$, $\forall x\in\ccalX$. $\hfill\Box$
	\end{definition}
	
	The probabilistic map $P_L$ provides a means to model dynamic and uncertain environments. Hereafter, we assume that the PL-MDP $\mathfrak{M}$ is fully observable, i.e., at any time/stage $t$ the current state, denoted by $x^t$, and the observations in state $x^t$, denoted by $\ell^t\in 2^\mathcal{AP}$, are known. 
	
	At any stage $T\geq0$ we define the robot's past path as $X_T=x_0x_1\dots x_T$, the past sequence of observed labels as $L_T=\ell_0\ell_1\dots \ell_T$, where $\ell_t\in 2^\mathcal{AP}$ and the past sequence of control actions $\ccalA_T=a_0a_1\dots a_{T-1}$, where $a_t\in\ccalA(x_t)$. These three sequences can be composed into a complete past run, defined as $R_T=x_0\ell_0 a_0 x_1\ell_1 a_1 \dots x_T\ell_T$. We denote by $\ccalX_T$, $\ccalL_T$, and $\ccalR_T$ the set of all possible sequences $X_T$, $L_T$ and $R_T$, respectively. 

	
	The goal of the robot is accomplish a task expressed as an LTL formula. LTL is a formal language that comprises a set of atomic propositions $\mathcal{AP}$, the Boolean operators, i.e., conjunction $\wedge$ and negation $\neg$, and two temporal operators, next $\bigcirc$ and until $\cup$. LTL formulas over a set $\mathcal{AP}$ can be constructed based on the following grammar: $$\phi::=\text{true}~|~\pi~|~\phi_1\wedge\phi_2~|~\neg\phi~|~\bigcirc\phi~|~\phi_1~\cup~\phi_2,$$ where $\pi\in\mathcal{AP}$. The other Boolean and temporal operators, e.g., \textit{always} $\square$, have their standard syntax and meaning. An infinite \textit{word} $\sigma$ over the alphabet $2^{\mathcal{AP}}$ is defined as an infinite sequence  $\sigma=\pi_0\pi_1\pi_2\dots\in (2^{\mathcal{AP}})^{\omega}$, where $\omega$ denotes infinite repetition and $\pi_k\in2^{\mathcal{AP}}$, $\forall k\in\mathbb{N}$. The language $\left\{\sigma\in (2^{\mathcal{AP}})^{\omega}|\sigma\models\phi\right\}$ is defined as the set of words that satisfy the LTL formula $\phi$, where $\models\subseteq (2^{\mathcal{AP}})^{\omega}\times\phi$ is the satisfaction relation \cite{pnueli}.
	
	In what follows, we define the probability that a stationary policy for $\mathfrak{M}$ satisfies the assigned LTL specification. Specifically, a stationary policy $\boldsymbol\xi$ for $\mathfrak{M}$ is defined as $\boldsymbol\xi=\xi_0\xi_1\dots$, where $\xi_t:\ccalX\times\ccalA\rightarrow[0,1]$. Given a stationary policy $\boldsymbol\xi$, the probability measure $\mathbb{P}_\mathfrak{M}^{\boldsymbol\xi}$, defined on the smallest $\sigma$-algebra over $\ccalR_{\infty}$, 
	is the unique measure defined as $\mathbb{P}_\mathfrak{M}^{\boldsymbol\xi}=\prod_{t=0}^T P_C(x_t,a_t,x_{t+1})P_L(x_t,\ell_t)\xi_t(x_t,a_t),$
	%
	where $\xi_t(x_t,a_t)$ denotes the probability that at time $t$ the action $a_t$ will be selected given the current state $x_t$ \cite{baier2008principles,puterman}.
	We then define the probability of $\mathfrak{M}$ satisfying $\phi$ under policy $\boldsymbol\xi$ as \cite{baier2008principles,clarke1999model}
	\begin{equation}\label{eq:probPhi}
	\mathbb{P}_\mathfrak{M}^{\boldsymbol\xi}(\phi)=\mathbb{P}_\mathfrak{M}^{\boldsymbol\xi}(\ccalR_{\infty}:\ccalL_{\infty}\models\phi), 
	\end{equation}
	
	The problem we address in this paper is summarized as follows.
	\begin{problem}\label{pr:pr1}
		Given a PL-MDP $\mathfrak{M}$ with unknown transition probabilities, unknown label mapping, unknown underlying graph structure, and a task specification captured by an LTL formula $\phi$, synthesize a deterministic stationary control policy $\boldsymbol\xi^*$ that maximizes the probability of satisfying $\phi$ captured in \eqref{eq:probPhi}, i.e., $\boldsymbol\xi^*=\argmax_{\boldsymbol\xi}\mathbb{P}_\mathfrak{M}^{\boldsymbol\xi}(\phi)$.\footnote{The fact that the graph structure is unknown implies that we do not know which transition probabilities are equal to zero. As a result, relevant approaches that require the structure of the MDP, as e.g., \cite{wang2015temporal} cannot be applied.}  
		$\hfill\Box$
	\end{problem}

	\section{A New Learning-for-Planning Algorithm}\label{sec:solution}
	
	In this section, we first discuss how to translate the LTL formula into an LDBA $\mathfrak{A}$ (see Section \ref{sec:LDBA}). 
	Then, we define the product MDP $\mathfrak{P}$, constructed by composing the PL-MDP $\mathfrak{M}$ and the LDBA $\mathfrak{A}$ that expresses $\phi$ (see Section \ref{sec:PMDP}). 
	Next, we assign rewards to the product MDP transitions based on the accepting condition of the LDBA $\mathfrak{A}$. 
	As we show later, this allows us to synthesize a policy $\boldsymbol\mu^*$ for $\mathfrak{P}$ that maximizes the probability of satisfying the acceptance conditions of the LDBA. 
	The projection of the obtained policy $\boldsymbol\mu^*$ over model $\mathfrak{M}$ results in a policy $\boldsymbol\xi^*$ that solves Problem~\ref{pr:pr1} (Section~\ref{sec:mPMDP}).

	\subsection{Translating LTL into an LDBA}\label{sec:LDBA}
	An LTL formula $\phi$ can be translated into an automaton, namely a finite-state machine that can express the set of words that satisfy $\phi$. 
	Conventional probabilistic model checking methods translate LTL specifications into DRAs, which are then composed with the PL-MDP, giving rise to a product MDP. 
	Nevertheless, it is known that this conversion results, in the worst case, in automata that are doubly exponential in the size of the original LTL formula \cite{dra4}. By contrast, in this paper we propose to express the given LTL property as an LDBA, which results in a much more succinct automaton~\cite{sickert,sickert2}. This is the key to the \textit{reduction of the state-space that needs to be explored}; see also Section \ref{sec:sim}. 
	
	Before defining the LDBA, we first need to define the Generalized B\"uchi Automaton (GBA).
	\begin{definition}[Generalized B\"uchi Automaton \cite{baier2008principles}]
		A GBA $\mathfrak{A} = (\ccalQ, q_0, \Sigma, \ccalF, \delta)$ is a structure where $\ccalQ$ is a finite set of states, $q_0\in\ccalQ$ is the initial state, $\Sigma=2^{\mathcal{AP}}$ is a finite alphabet, $\ccalF = {\ccalF_1,\dots, \ccalF_f }$ is the set of accepting conditions where $\ccalF_j \subset\ccalQ$, $1\leq j \leq f$, and $\delta:\ccalQ\times\Sigma\rightarrow 2^{\ccalQ}$ is a transition relation. $\hfill\Box$ 
	\end{definition}
	
	An infinite run $\rho$ of $\mathfrak{A}$ over an infinite word $\sigma=\pi_0\pi_1\pi_2\dots\in\Sigma^{\omega}$, $\pi_k\in\Sigma=2^{\mathcal{AP}}$ $\forall k\in\mathbb{N}$, is an infinite sequence of states $q_k\in\ccalQ$, i.e.,  $\rho=q_0q_1\dots q_k\dots$, such that $q_{k+1}\in\delta(q_k,\pi_k)$. The infinite run $\rho$ is called \textit{accepting} (and the respective word $\sigma$ is accepted by the GBA) if $\texttt{Inf}(\rho)\cap\ccalF_j\neq\emptyset,\forall j\in\set{1,\dots,f},$ %
	where $\texttt{Inf}(\rho)$ is the set of states that are visited infinitely often by $\rho$. 
	
	\begin{definition}[Limit Deterministic B\"uchi Automaton \cite{sickert}]\label{ldbadef}
		A GBA $\mathfrak{A} = (\ccalQ, q_0, \Sigma, \ccalF, \delta)$ is {\it limit deterministic} if
		$\ccalQ$ can be partitioned into two disjoint sets $\ccalQ = \ccalQ_N \cup \ccalQ_D$, so that (i) $\delta(q,\pi)\subset\ccalQ_D$ and $|\delta(q,\pi)|=1$, for every state $q\in\ccalQ_D$ and $\pi\in\Sigma$; and (ii) for every $\ccalF_j\in\ccalF$, it holds that $\ccalF_j\subset\ccalQ_D$
		and there are $\varepsilon$-transitions from $\ccalQ_N$ to $\ccalQ_D$. $\hfill\Box$ 
	\end{definition}
	
	An $\varepsilon$-transition allows the automaton to change its state without reading any specific input. In practice, the $\varepsilon$-transitions between $ \mathcal{Q}_N $ and $ \mathcal{Q}_D $ reflect the ``guess'' on reaching $ \mathcal{Q}_D $: accordingly, if after an $\varepsilon$-transition the associated labels in the accepting set of the automaton cannot be read, or if the accepting states cannot be visited, then the guess is deemed to be wrong, and the trace is disregarded and is not accepted by the automaton. However, if the trace is accepting, then the trace will stay in $\mathcal{Q}_D$ ever after, i.e. $\mathcal{Q}_D$ is invariant. 
	
	\begin{definition}[Non-accepting Sink Component]\label{sink}
		A non-accepting sink component in an LDBA $\mathfrak{A}$ is a directed graph induced by a set of states $\ccalQ_{sink} \subset\mathcal{Q}$ such that (1) is strongly connected, (2) does not include all accepting sets $\ccalF_j,~j=1,...,f $, and (3) there exist no other strongly connected set $ \ccalQ' \subset \mathcal{Q},~\ccalQ'\neq \ccalQ_{sink} $ that $ \ccalQ_{sink} \subset \ccalQ' $. We denote the union set of all non-accepting sink components as $\mathcal{Q}_{sinks}$. $\hfill\Box$ 
	\end{definition}

	\subsection{Product MDP}\label{sec:PMDP} 
	
	Given the PL-MDP $\mathfrak{M}$ and the LDBA  $\mathfrak{A}$, we define the product MDP $\mathfrak{P}=\mathfrak{M} \times \mathfrak{A}$ as follows.
	\begin{definition}[Product MDP]\label{def:prodMDP}
		Given a PL-MDP $\mathfrak{M} \allowbreak = (\ccalX\allowbreak, x_0\allowbreak,  \ccalA\allowbreak, P_C\allowbreak, \mathcal{AP}\allowbreak, P_L)$ and an LDBA $\mathfrak{A} = (\ccalQ, q_0, \Sigma, \ccalF, \delta)$, we define the product MDP $\mathfrak{P}=\mathfrak{M} \times \mathfrak{A}$ as  $\mathfrak{P}\allowbreak = (\mathcal{S}\allowbreak, {s}_0\allowbreak, \ccalA_\mathfrak{P}\allowbreak, P_{\mathfrak{P}}\allowbreak, \mathcal{F}_\mathfrak{P})$, where 
		(i) $\mathcal{S} = \ccalX \times 2^{\mathcal{AP}}\times \ccalQ$ is the set of states, so that $s=(x,\ell,q)\in\ccalS$, $x\in\ccalX$, $\ell\in 2^{\mathcal{AP}}$, and $q\in\ccalQ$ ; 
		(ii) ${s_0} = (x_0, \ell_0,q_0)$ is the initial state; 
		(iii) $\mathcal{A}_\mathfrak{P}$ is the set of actions inherited from the MDP, so that $\mathcal{A}_\mathfrak{P} (s) = \ccalA(x)$, where $s=(x,\ell,q)$;  
		(iv) $P_\mathfrak{P}:\ccalS\times\ccalA\times\ccalS:[0,1]$ is the transition probability function, so that 
		\begin{equation}
		P_\mathfrak{P}([x,\ell,q],a,[x',\ell',q'])=P_{C}(x,u,x')P_L(x',\ell'),
		\end{equation}
		where $[x,\ell,q]\in\ccalS$, $[x',\ell',q']\in\ccalS$, $a\in\ccalA(x)$ and $q'=\delta(q,\ell')$; 
		(v) $\mathcal{F}_\mathfrak{P} = \{(\mathcal{F}_j^\mathfrak{P}), j=1,\dots,f\}$ is the set of accepting states, where $\mathcal{F}_j^\mathfrak{P} =  \ccalX \times 2^{\mathcal{AP}}\times \mathcal{F}_j$.
		In order to handle $\varepsilon$-transitions in the constructed LDBA we have to add the following modifications to the standard definition of the product MDP \cite{sickert2}. First, for every $\varepsilon$-transition to a state $q'\in\ccalQ$ we add an action $\varepsilon_{q'}$ in the product MDP, i.e., $\ccalA_\mathfrak{P}(s)= \ccalA_\mathfrak{P}(s)\cup \{\varepsilon_{q'}, s'=[x,\ell',q'], q'\in\ccalQ \}$. Second, the transition probabilities of $\varepsilon$-transitions are given by 
		\begin{equation}
		P_\mathfrak{P}(s,a,s')=
		\begin{cases}
		1,~\text{if}~(x=x')\wedge(\ell=\ell')\wedge(\delta(q,\varepsilon_{q'})=q')\\
		0,~\text{otherwise}, 
		\end{cases}
		\end{equation}
		where $s=(x,\ell,q)$ and $s'=(x',\ell',q')$. $\hfill\Box$
	\end{definition}

	Given  any policy $\boldsymbol\mu$ for $\mathfrak{P}$, we define an infinite run $\rho_{\mathfrak{P}}^{\boldsymbol\mu}$ of $\mathfrak{P}$ to be an infinite sequence of states of $\mathfrak{P}$, i.e., $\rho_{\mathfrak{P}}^{\boldsymbol\mu}=s_0s_1s_2\dots$, where $P_{\mathfrak{P}}(s_t,\boldsymbol\mu(s_t),s_{t+1})>0$. By definition of the accepting condition of the LDBA $\mathfrak{A}$, an infinite run  $\rho_{\mathfrak{P}}^{\boldsymbol\mu}$ is accepting, i.e., $\boldsymbol\mu$ satisfies $\phi$ with a non-zero probability (denoted by $\boldsymbol\mu\models\phi$), if $\texttt{Inf}(\rho_{\mathfrak{P}}^{\boldsymbol\mu})\cap\ccalF^\mathfrak{P}_j\neq\emptyset$, $\forall j\in\set{1,\dots,f}$.

	In what follows, we design a synchronous reward function based on the accepting condition of the LDBA so that maximization of the expected accumulated reward implies maximization of the satisfaction probability. Specifically, we generate a control policy $\boldsymbol\mu^*$ that maximizes the probability of (i) reaching the states of $\ccalF_\mathfrak{P}$ from $s_0$ and (ii) the probability that each accepting set $\ccalF^\mathfrak{P}_j$ will be visited infinitely often. 
	
	\subsection{Construction of the Reward Function}\label{sec:mPMDP}
	
	To synthesize a policy that maximizes the probability of satisfying $\phi$, we construct a synchronous reward function for the product MDP. The main idea is that (i) visiting a set $\ccalF_j$, $1\leq j\leq f$ yields a positive reward $r>0$; and (ii) revisiting the same set $\ccalF_j$ returns zero reward until all other sets $\ccalF_k$, $k\neq j$ are also visited; (iii) the rest of the transitions have zero rewards. Intuitively, this reward shaping strategy motivates the agent to visit \textit{all} accepting sets $\ccalF_j$ of the LDBA infinitely often, as required by the acceptance condition of the LDBA; see also Section \ref{sec:guarantees}.
	
	To formally present the proposed reward shaping method, we need first to introduce the 
	the \textit{accepting frontier set} $ \mathds{A} $ which is initialized as the family set
	\begin{equation}\label{eq:initA}
	\mathds{A}=\{F_k\}_{k=1}^{f}.
	\end{equation}
	This set is updated on-the-fly every time a set $\ccalF_j$ is visited as $\mathds{A}\leftarrow AF(q,\mathds{A})$ 
	where $AF(q,\mathds{A})$ is the \textit{accepting frontier function} defined as follows.
	
	\begin{definition}
		[Accepting Frontier Function] \label{frontier} Given an LDBA $\mathfrak{A} \allowbreak =( \mathcal{Q} \allowbreak, q_0\allowbreak,\Sigma\allowbreak,\allowbreak\mathcal{F},\allowbreak\delta)$, we define $ AF:\mathcal{Q}\times 2^{\mathcal{Q}}\rightarrow2^\mathcal{Q} $ as the accepting frontier function, 
		which executes the following operation over any given set $ \mathds{A}\in 2^{\mathcal{Q}}$: 	
		\[
		AF(q,\mathds{A} )=\left\{
		\begin{array}{l@{\hspace{0.2cm}:\hspace{0.2cm}}l}
		\mathds{A} \setminus \ccalF_j~~~ & (q \in \ccalF_j) \wedge (\mathds{A}\neq \ccalF_j)\\
		\\
		\{F_k\}_{k=1}^{f} \setminus \ccalF_j & (q \in \ccalF_j) \wedge (\mathds{A}=\ccalF_j).\hspace{0.65cm}\Box\\
		\end{array}
		\right. 	
		\]
	\end{definition}
	\medskip
	In words, given a state $ q\in \ccalF_j $ and the set $ \mathds{A} $, $ AF $ outputs a set containing the elements of $\mathds{A}$ minus those elements that are common with $ \ccalF_j$ (first case). However, if $ \mathds{A}=\ccalF_j $, then the output is the family set of all accepting sets of $\mathfrak{A}$ minus those elements that are common with $ \ccalF_j $, resulting in a reset of $\mathds{A}$ to \eqref{eq:initA} minus those elements that are common with $ \ccalF_j $ (second case). Intuitively, $\mathds{A}$ always contains those accepting sets that are needed to be visited at a given time and in this sense the reward function is synchronous with the LDBA accepting condition.

	Given the accepting frontier set $\mathds{A}$, we define the following reward function
	\begin{equation}\label{def:prodRewMDPi}
	\begin{aligned}
	R(s,a) = \left\{
	\begin{array}{lr}
	r & \text{if}~q'~\in \mathds{A},~{s}'=(x',\ell',q'),\\
	0 & \text{otherwise}.
	\end{array}
	\right.
	\end{aligned}
	\end{equation} 
	
	In \eqref{def:prodRewMDPi}, ${s}'$ is the state of the product MDP that is reached from state ${s}$ by taking action $a$, and $r>0$ is an arbitrary positive reward. In this way the agent is guided to visit all accepting sets $\ccalF_j$ infinitely often and, consequently, satisfy the given LTL property.  
	
	\begin{rem}\label{on_the_fly_remark}
		The initial and accepting components of the LDBA proposed in \cite{sickert} (as used in this paper) are both deterministic. By Definition \ref{ldbadef}, the discussed LDBA is indeed a limit-deterministic automaton, however notice that the obtained determinism within its initial part is stronger than that required in the definition of LDBA. Thanks to this feature of the LDBA structure, in our proposed algorithm there is no need to ``explicitly build'' the product MDP and to store all its states in memory. The automaton transitions can be executed on-the-fly, as the agent reads the labels of the MDP states. $\hfill\Box$
	\end{rem}
	
	Given $\mathfrak{P}$, we compute a \textit{stationary deterministic} policy $\boldsymbol\mu^*$, that maximizes the expected accumulated return, i.e.,
	
	\begin{equation}\label{eq:policy0}
	\boldsymbol\mu^*(s)=\arg\max\limits_{\boldsymbol\mu \in \mathcal{D}}~ {U}^{\boldsymbol\mu}(s),
	\end{equation}
	where $\mathcal{D}$ is the set of all stationary deterministic policies over $\mathcal{S}$, and 	
	\begin{equation}\label{eq:utility}
	{U}^{\boldsymbol\mu}(s)=\mathds{E}^{\boldsymbol\mu} [\sum\limits_{n=0}^{\infty} \gamma^n~ R(s_n,\boldsymbol\mu(s_n))|s_0=s],
	\end{equation}
	where $\mathds{E}^{\boldsymbol\mu} [\cdot]$ denotes the expected value given that the product MDP follows the policy $\boldsymbol\mu$ \cite{puterman}, $0\leq\gamma\leq 1$ is the discount factor, and $s_0,...,s_n$ is the sequence of states generated by policy $\boldsymbol\mu$ up to time step $n$, initialized at $s_0 = s$. Note that the optimal policy is stationary as shown in the following result.
	
	\begin{theorem}[\!\!\cite{puterman}]
		In any finite-state MDP, such as $\mathfrak{P}$, if there exists an optimal policy, then that policy is stationary and deterministic. $\hfill\Box$
	\end{theorem}
	

	
	In order to construct $\boldsymbol\mu^*$, we employ episodic Q-learning (QL), a model-free RL scheme described in Algorithm~\ref{alg:stationary}.\footnote{Note that any other off-the-shelf model-free RL algorithm can also be used within Algorithm~\ref{alg:stationary}, including any variant of the class of temporal difference learning algorithms~\cite{sutton}.} 
	Specifically, Algorithm \ref{alg:stationary} requires as inputs (i) the LDBA $\mathfrak{A}$, (ii) the reward function $R$ defined in \eqref{def:prodRewMDPi}, and (iii) the hyper-parameters of the learning algorithm. 
	
	Observe that in Algorithm \ref{alg:stationary}, we use an action-value function $Q:\mathcal{S}\times\mathcal{A}_\mathfrak{P}\rightarrow\mathbb{R}$ to evaluate $\boldsymbol\mu$ instead of $U^{\boldsymbol\mu}(s)$, since the MDP $\mathfrak{P}$ is unknown. The action-value function $Q(s,a)$ can be initialized arbitrarily. Note that $U^{\boldsymbol\mu}(s)=\max_{a \in \mathcal{A}_\mathfrak{P}}Q(s,a)$. Also, we define a function $C:\mathcal{S}\times\mathcal{A}_\mathfrak{P}\rightarrow\mathbb{N}$ that counts the number of times that action $a$ has been taken at state $s$. The policy $\boldsymbol\mu$ is selected to be an $\epsilon$-greedy policy, which means that with probability $1-\epsilon$, the greedy action $\argmax_{a\in\mathcal{A}_\mathfrak{P}}Q(s,a)$ is taken, and with probability $\epsilon$ a random action $a$ is selected. Every episode terminates when the current state of the automaton gets inside $\mathcal{Q}_{sinks}$ (Definition \ref{sink}) or when the iteration number in the episode reaches a certain threshold $\tau$. Note that it holds that $\boldsymbol\mu$ asymptotically converges to the optimal greedy policy $\boldsymbol\mu^*=\argmax_{a\in\mathcal{A}_\mathfrak{P}} Q^*(s,a)$:
	where $Q^*$ is the optimal $Q$ function. Further, $Q(s,\boldsymbol\mu^*(s))=U^{\boldsymbol\mu^*}(s)=V^*({s})$, where $V^*({s})$ is the optimal value function that could have been computed via Dynamic Programming (DP) if the MDP was fully known \cite{sutton,reachability_in_hybrid,cont_pctl}. Projection of $\boldsymbol\mu^*$ onto the state-space of the PL-MDP, yields the finite-memory policy $\boldsymbol\xi^*$ that solves Problem \ref{pr:pr1}.
	
	\begin{algorithm2e}[!t]
		\DontPrintSemicolon
		\SetKw{return}{return}
		\SetKwRepeat{Do}{do}{while}
		\SetKwFunction{terminate}{episode$\_$terminate}
		\SetKwFor{terminatedef}{episode$\_$terminate()}{}{}
		\SetKwData{conflict}{conflict}
		\SetKwData{safe}{safe}
		\SetKwData{sat}{sat}
		\SetKwData{unsafe}{unsafe}
		\SetKwData{unknown}{unknown}
		\SetKwData{true}{true}
		\SetKwData{false}{false}
		\SetKwInOut{Input}{input}
		\SetKwInOut{Output}{output}
		\SetKwFor{Loop}{Loop}{}{}
		\SetKw{KwNot}{not}
		\begin{small}
			\Input {Reward function $R$, LDBA $\mathfrak{A}$,$\gamma$,$\tau$}
			\Output {$\boldsymbol\mu^*$}
			Initialize $C(s,a)=0$, $Q(s,a),~\forall s \in \mathcal{S},~\forall a \in \mathcal{A}_\mathfrak{P}$\; 
			\hspace{11mm} $\mathds{A}=\bigcup_{k=1}^{f} F_k $\;
			\hspace{11mm} $episode$-$number:=0$, $iteration$-$number:=0$\;
			\While{$Q$ is not converged}
			{
				$episode$-$number++$\;
				${s}_\text{cur}={s}_0$\;
				$\epsilon=1/(episode$-$number)$\;
				\While{$(q\not\in\mathcal{Q}_{sinks})\wedge(iteration$-$number<\tau)$}
				{
					$iteration$-$number++$\;
					Set $a_{\text{cur}}=\argmax_{a\in\mathcal{A}_\mathfrak{P}}Q(s,a)$ with probability $1-\epsilon$ and set $a_{\text{cur}}$ as a random action in $\mathcal{A}_\mathfrak{P}$ with probability $\epsilon$\; 
					Execute $a_{\text{cur}}$ and observe ${s}_{\text{next}}=(x_{next},\ell_{next},q_{next})$, and $R({s}_\text{cur},a_{\text{cur}})$\;
					\uIf{$R({s}_\text{cur},a_{\text{cur}})>0$}
					{
						$\mathds{A}=AF(q_\text{next},\mathds{A}),$\;
					}
					$C({s}_\text{cur},a_{\text{cur}})++$\;
					$Q({s}_\text{cur},a_{\text{cur}})= Q({s}_\text{cur},a_{\text{cur}})+(1/C({s}_\text{cur},a_{\text{cur}}))[R({s}_\text{cur},a_{\text{cur}})-Q({s}_\text{cur},a_{\text{cur}})+\gamma \max_{a'}({s}_{\text{next}},a'))]$\;
					${s}_\text{cur}={s}_{\text{next}}$
				}
			}
		\end{small}
		\caption{RL for LTL objective}
		\label{alg:stationary}
	\end{algorithm2e}

	\section{Analysis of the Algorithm}\label{sec:guarantees}
	In this section, we show that the policy $\boldsymbol{\mu}^*$ generated by Algorithm \ref{alg:stationary} maximizes \eqref{eq:probPhi}, i.e., the probability of satisfying the property $\phi$. Furthermore, we show that, unlike existing approaches, our algorithm can produce the best available policy if the property cannot be satisfied. To prove these claims, we need to show the following results. All proofs are presented in the Appendix.
	First, we show that the accepting frontier set $\mathds{A}$ is time-invariant. This is needed to ensure that the LTL formula is satisfied over the product MDP by a stationary policy.
	
	\begin{prop}\label{prop:timeInv}
		For an LTL formula $ \phi $ and its associated LDBA $\mathfrak{A}=\allowbreak(\mathcal{Q}\allowbreak,q_0\allowbreak,\Sigma\allowbreak, \mathcal{F}\allowbreak, \delta\allowbreak)$, the accepting frontier set $ \mathds{A} $ is time-invariant at each state of $\mathfrak{A}$. $\hfill\Box$ 
	\end{prop}
	
	As stated earlier, since QL is proved to converge to the optimal Q-function~\cite{sutton}, it can synthesize an optimal policy with respect to the given reward function. The following result shows that the optimal policy produced by Algorithm~\ref{alg:stationary} satisfies the given LTL property. 
	
	\begin{theorem}\label{thm1}
		Assume that there exists at least one deterministic stationary policy in $\mathfrak{P}$ whose traces satisfy the property $\phi$ with positive probability. Then the traces of the optimal policy $\boldsymbol{\mu}^*$ defined in \eqref{eq:policy0} satisfy $\phi$ with positive probability, as well.
	\end{theorem}
	
	Next we show that $\boldsymbol\mu^*$ and subsequently its projection $\boldsymbol\xi^*$ maximize the satisfaction probability.
	
	\begin{theorem}\label{thm2}
		If an LTL property $\phi$ is satisfiable by the PL-MDP~$\mathfrak{M}$, then the optimal policy $\boldsymbol{\mu}^*$ that maximizes the expected accumulated reward, as defined in \eqref{eq:policy0}, maximizes the probability of satisfying $\phi$, defined in \eqref{eq:probPhi}, as well. $\hfill\Box$
	\end{theorem}
	
	Next, we show that if there does not exist a policy that satisfies the LTL property $\phi$, Algorithm \ref{alg:stationary} will find the policy that is the closest one to property satisfaction. To this end, we first introduce the notion of \textit{closeness to satisfaction}.
	
	\begin{definition}[Closeness to Satisfaction]\label{def:closesat}
		Assume that two policies $ {\boldsymbol\mu}_1 $ and $ {\boldsymbol\mu}_2 $ do not satisfy the property $ \phi $. Consequently, there are accepting sets in the automaton that have no intersection with runs of the induced Markov chains $ \mathfrak{P}^{{\boldsymbol\mu}_1} $ and $ \mathfrak{P}^{{\boldsymbol\mu}_2} $. 
		The policy $ {\boldsymbol\mu}_1 $ is closer to satisfying the property if runs of $ \mathfrak{P}^{{\boldsymbol\mu}_1} $ have more intersections with accepting sets of the automaton than runs of $ \mathfrak{P}^{{\boldsymbol\mu}_2} $. $\hfill\Box$ 
	\end{definition}
	
	\begin{cor}\label{cor2}
		If there does not exist a policy in the PL-MDP~$\mathfrak{M}$ that satisfies the property $ \phi $, then proposed algorithm yields a policy that is closest to satisfying $\phi$. $\hfill\Box$ 
	\end{cor}
	
	\section{Experiments}\label{sec:sim}
	
	\begin{figure}[t]
		\centering
		\includegraphics[width=0.7\linewidth]{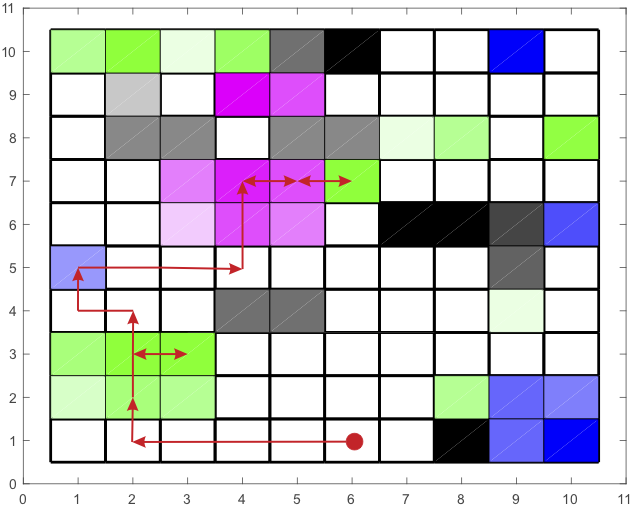}
		\caption{PL-MDP that models the interaction of the robot with the environment. The color of each region (square) corresponds to the probability that some event can be observed there. Specifically, gray, magenta, blue, and green mean that there is a non-zero probability of an obstacle (${\text{obs}}$), a user (${\text{user}}$), target~1 (${\text{target1}}$), and target~2 ${\text{target2}}$. Higher intensity indicates a higher probability. The red trajectory represents a sample path of the robot with the optimal control strategy~$\boldsymbol\xi^*$ for the first case study. The red dot is the initial location of the robot.}
		\label{fig:env}
	\end{figure}
	
	\begin{figure}[t]
		\centering
		\includegraphics[width=0.7\linewidth]{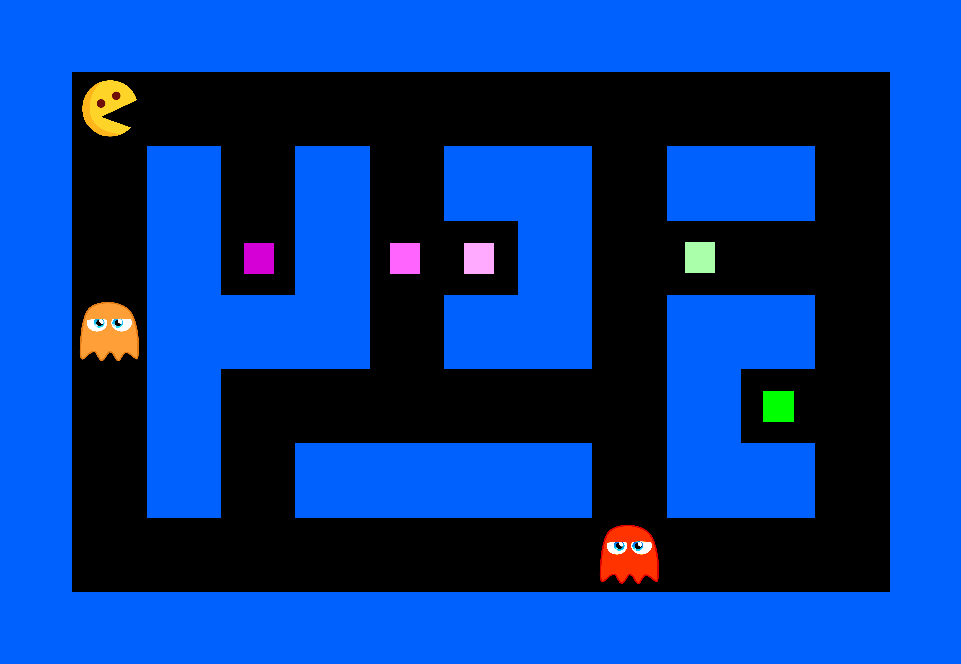}
		\caption{Initial condition in Pacman environment. The magenta square is labeled $\text{food1}$ and the green one $\text{food2}$. The color intensity of each square corresponds to the probability of the food being observed. The state of being caught by a ghost is labeled $\text{ghost}$ and the rest of the state space $\text{neutral}$.}
		\label{fig:pacman_initial}
	\end{figure}
	
	\begin{figure*}[t]
		\centering
		\subfigure[Case Study I]{
			\label{fig:uamec}
			\includegraphics[width=0.3\linewidth]{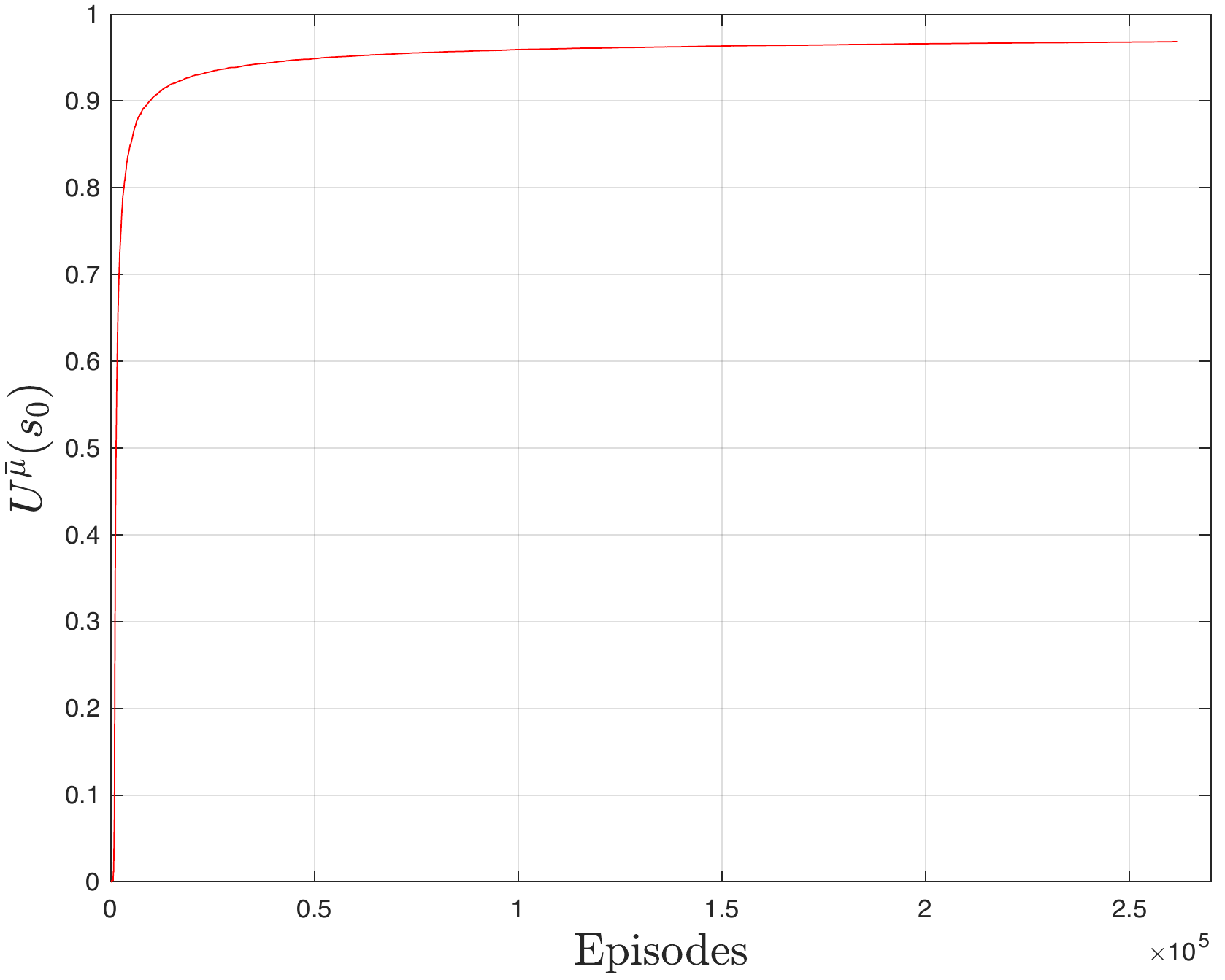}}\hspace*{0.5cm}
		\subfigure[Case Study II]{
			\label{fig:uNoAMEC}
			\includegraphics[width=0.3\linewidth]{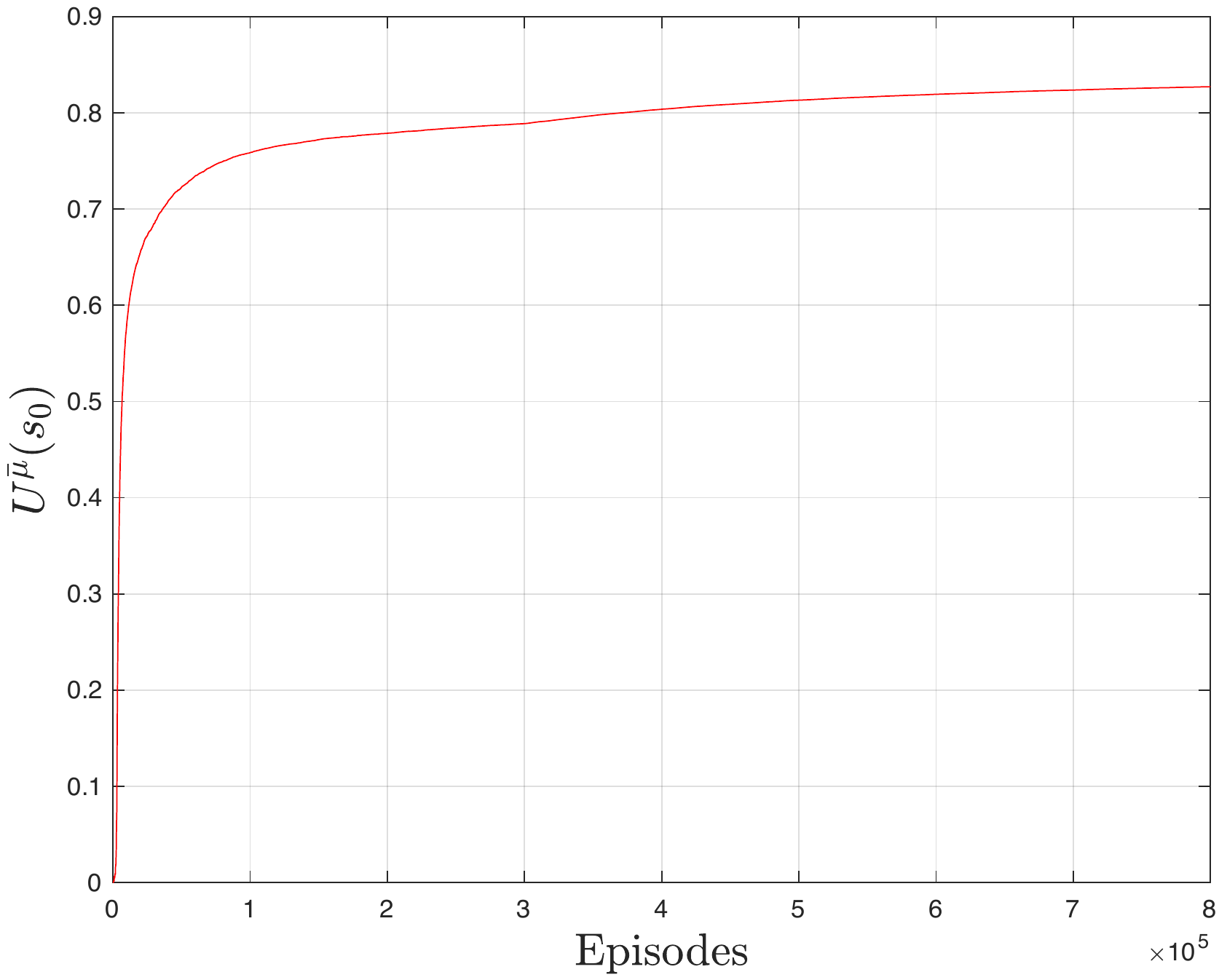}}\hspace*{0.5cm}
		\subfigure[Case Study III]{
			\label{fig:pacman_result}
			\includegraphics[width=0.3\linewidth]{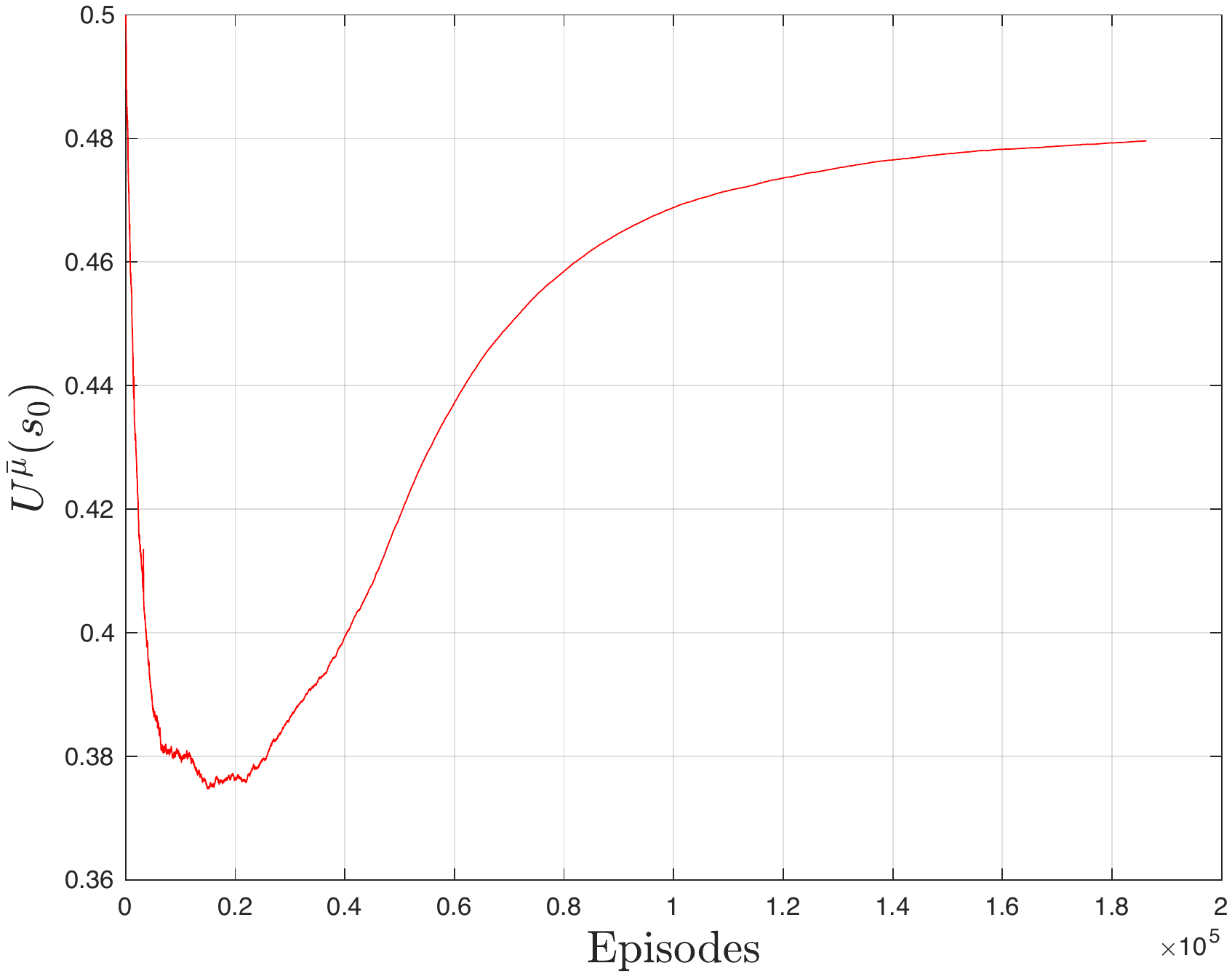}}
		\caption{\hosein{Illustration of the evolution of $U^{\bar{\boldsymbol\mu}}(s_0)$ with respect to episodes. $\bar{\boldsymbol\mu}$ denotes the $\epsilon$-greedy policy which converges to the optimal greedy policy $\boldsymbol\mu^*$. Videos of Pacman winning the game can be found in \cite{website}}}
		\label{fig:results}
	\end{figure*}
	
	In this section we present three case studies, implemented on MATLAB R2016a on a computer with an Intel Xeon CPU at 2.93\,GHz and 4\,GB RAM. In the first two experiments, the environment is represented as a $10\times 10$ discrete grid world, as illustrated in Figure~\ref{fig:env}. The third case study is an adaptation of the well-known Atari game Pacman (Figure~\ref{fig:pacman_initial}), which is initialized in a configuration that is quite hard for the agent to solve. 
	
	The first case study pertains to a temporal logic planning problem in a dynamic and unknown environment with AMECs, while the second one does not admit AMECs. Note that the majority of existing algorithms fail to provide a control policy when AMECs do not exist \cite{ding2014optimal,fu2014probably,wang2015temporal}, or result in control policies without satisfaction guarantees  \cite{dorsa}. 
	
	The LTL formula considered in the first two case studies is the following:  
	\begin{align}\label{phi_1}
	\begin{aligned}
	\phi_1=&\Diamond ({\text{target1}})\wedge\square\Diamond ({\text{target2}})\wedge\square\Diamond ({\text{user}})\wedge\\
	&(\neg {\text{user}} \cup {\text{target2}})\wedge \square (\neg {\text{obs}}).
	\end{aligned}
	\end{align}
	
	%
	%
	In words, this LTL formula requires the robot  to (i)~eventually visit target~1 (expressed as  $\Diamond{\text{target1}}$); (ii)~visit target~$2$ infinitely often and take a picture of it 
	($\square\Diamond{\text{target2}}$); (iii)~visit a user infinitely often where, say, the collected pictures are uploaded (captured by $\square\Diamond{\text{user}}$); (iv)~avoid visiting the user until a picture of target~$2$ has been taken; and (v)~always avoid obstacles (captured by $\square (\neg {\text{obs}}$). 
	
	The LTL formula~\eqref{phi_1} can be expressed as a DRA with $11$ states. On the other hand, a corresponding LDBA has $5$ states (fewer, as expected), which results in a significant reduction of the state space that needs to be explored.  
	
	The interaction of the robot with the environment is modeled by a PL-MDP $\mathfrak{M}$ with $100$ states and $10$ actions per state. The actions space is $\{\mathit{Up}, \mathit{Right}, \mathit{Down}, \mathit{Left}, \mathit{None}\}\times\{\mathit{Take~picture}, \mathit{Do~not~take~picture}\}$. 
	We assume that the targets and the user are dynamic, i.e., their location in the environment varies probabilistically. Specifically, their presence in a given region $x\in\ccalX$ is determined by the unknown function $P_L$ from Definition~\ref{def:labelMDP} (Figure \ref{fig:env}). 
	
	The LTL formula specifying the task for Pacman (third case study) is:
	\begin{align}\label{phi_2}
	\begin{aligned}
	\phi_2=&\lozenge [ (\text{food1} \wedge \lozenge \text{food2}) \vee (\text{food2} \wedge \lozenge \text{food1})]  \wedge\\ &\square(\neg \text{ghost}).
	\end{aligned}
	\end{align}
	Intuitively, the agent is tasked with (i) eventually eating $\text{food1}$ and then $\text{food2}$ (or vice versa), while (ii) avoiding any contact with the ghosts. This LTL formula corresponds to a DRA with $5$ states and to an LDBA with $4$ states. 
	The agent can execute $5$ actions per state $\{\mathit{Up}, \mathit{Right}, \mathit{Down}, \mathit{Left}, \mathit{None}\}$ and if the agent hits a wall by taking an action it remains in the previous location. 
	The ghosts dynamics are stochastic: with a probability $ p_g=0.9 $ each ghost chases the Pacman (often referred to as ``chase mode''), and with its complement it executes a random action (``scatter mode'').  
	
	In the first case study, we assume that there is no uncertainty in the robot actions. In this case, it can be verified that AMECs exist. Figure~\ref{fig:uamec} illustrates the evolution of $U^{\bar{\boldsymbol\mu}}(s_0)$ over $260000$ episodes, where $\bar{\boldsymbol\mu}$ denotes the $\epsilon$-greedy policy. The optimal policy was constructed in approximately $30$ minutes. 
	A sample path of the robot with the projection of optimal control strategy $\boldsymbol\mu^*$ onto $\mathcal{X}$, i.e. policy $\boldsymbol\xi^*$, is given in Figure~\ref{fig:env} (red path).
	
	
	
	
	
	In the second case study, we assume 
	that the robot is equipped with a noisy controller and, therefore, it can execute the desired action with probability $0.8$, 
	whereas a random action among the other available ones is taken with a probability of $0.2$. 
	In this case, it can be verified that AMECs do not exist. Intuitively, the reason why AMECs do not exist is that there is always a non-zero probability with which the robot will hit an obstacle while it travels between the access point and target $2$ and, therefore, it will violate $\phi$.
	Figure~\ref{fig:uNoAMEC} shows the evolution of $U^{\bar{\boldsymbol\mu}}(s_0)$ over $800000$ episodes for the $\epsilon$-greedy policy. 
	The optimal policy was synthesized in approximately $2$ hours.
	
	
	

	
	In the third experiment, there is no uncertainty in the execution of actions, namely the motion of the Pacman agent is deterministic. Figure \ref{fig:pacman_result} shows the evolution of $U^{\bar{\boldsymbol\mu}}(s_0)$ over 186000 episodes where $\bar{\boldsymbol\mu}$ denotes the $\epsilon$-greedy policy. On the other hand, the use of standard Q-learning (without LTL guidance) would require either to construct a history-dependent reward for the PL-MDP $\mathfrak{M}$ as a proxy for the considered LTL property, which is very challenging for complex LTL formulas, 
	or to perform exhaustive state-space search with static rewards, which is evidently quite wasteful and failed to generate an optimal policy in our experiments.
	
	Note that given the policy $\boldsymbol{\xi}^*$ for the PL-MDP, probabilistic model checkers, such as PRISM \cite{kwiatkowska2002prism}, or standard Dynamic Programming methods can be employed to compute the probability of satisfying $\phi$. For instance, for the first case study, the synthesized policy satisfies $\phi$ with probability $1$, while for the second case study, the satisfaction probability is $0$, since AMECs do not exist. For the same reason, even if the transition probabilities of the PL-MDP are known, PRISM could not generate a policy for the second case study. Nevertheless, the proposed algorithm can synthesize the closest-to-satisfaction policy, as shown in Corollary \ref{cor2}.
	
	\section{Conclusions}
	
	In this paper we have proposed a model-free reinforcement learning (RL) algorithm to synthesize control policies that maximize the probability of satisfying high-level control objectives captured by LTL formulas. The interaction of the agent with the environment has been captured by an unknown probabilistically-labeled Markov Decision Process (MDP). We have shown that the proposed RL algorithm produces a policy that maximizes the satisfaction probability. We have also shown that even if the assigned specification cannot be satisfied, the proposed algorithm synthesizes the best possible policy. We have provided evidence via numerical experiments on the efficiency of the proposed method. 
	
	\clearpage
	
	\bibliographystyle{IEEEtran}
	\bibliography{YK_bib}
	
	\clearpage
	
	\appendix
	
	\begin{definition}\label{g_monitor}
		Given an LTL property $ \phi $ and a set $\mathcal{G}$ of  G-subformulas, i.e., formulas in the form $ \square(\cdot)$, we define $ \phi[\mathcal{G}] $ to be the resulting formula when we substitute $\mathit{true}$ for every G-subformula in $ \mathcal{G} $ and $ \neg \mathit{true} $ for other G-subformulas of~$\phi$.  
	\end{definition}
	
	\subsection{Proof of Proposition \ref{prop:timeInv}}
	Let $ \mathcal{G}=\{\square\zeta_1,...,\square\zeta_f\} $ be the set of all G-subformulas of~$\phi$. Since elements of $ \mathcal{G} $ are subformulas of $ \phi $ we can assume an ordering over $ \mathcal{G} $ so that if $ \square\zeta_i $ is a subformula of $ \square\zeta_j $ then $ j>i $. The accepting component of LDBA $ \mathcal{Q}_D $ is a product of $ f $ DBAs $ \{\mathfrak{D}_1,....,\mathfrak{D}_f\} $ (called G-monitors) such that each $ \mathfrak{D}_i = (\mathcal{Q}_i,{q_i}_0,\Sigma,F_i,\delta_i)$ expresses $ \square\zeta_i[\mathcal{G}] $ where $\mathcal{Q}_i $ is the state space of the $ i $-th G-monitor, $ \Sigma=2^{\mathcal{AP}} $, and $ \delta_i:\mathcal{Q}_i\times\Sigma\rightarrow\mathcal{Q}_i $ \cite{sickert}. Note that $ \zeta_i[\mathcal{G}] $ has no G-subformulas. The states of the G-monitor $ \mathfrak{D}_i $ are pairs of formulas where at each state, the first checks if the run satisfies $ \square\zeta_i[\mathcal{G}] $, while the second puts the next G-subformula in the ordering of $ \mathcal{G} $ on hold. However, all the previous G-subformulas have been checked already and is replaced by $true$ in $\square\zeta_i[\mathcal{G}]$. 
	The product of the G-monitors is a deterministic generalized B\"uchi automaton: 
	$\mathfrak{A}_{D}=(\allowbreak\mathcal{Q}_D,\allowbreak {q_D}_0,\allowbreak\Sigma, \allowbreak\mathcal{F}, \allowbreak\delta),$
	where $ \mathcal{Q}_D=\mathcal{Q}_1\times...\times\mathcal{Q}_f $, $ \Sigma=2^{\mathcal{AP}} $, $ \mathcal{F}=\{\ccalF_1,...,\ccalF_f\} $, and $ \delta=\delta_1\times...\times\delta_f $.
	As shown in \cite{sickert}, while a word $ w $ is being read by the accepting component of the LDBA, the set of G-subformulas that hold ``monotonically'' expands. If $ w \in \left\{\sigma\in (2^{\mathcal{AP}})^{\omega}|\sigma\models\phi\right\} $, then eventually all G-subformulas become true. 
	
	Assume that the current state of the automaton is $ q_D=(q_1,...,q_i,...,q_f) $ and the automaton is checking whether $ \square\zeta_i[\mathcal{G}] $ is satisfied or not, assuming that $ \square\zeta_{i-1} $ is already $true$, while putting $\square\zeta_{i+1}[\mathcal{G}]$ on hold. At this point, the accepting frontier set is $ \mathds{A}=\{\ccalF_{i},\ccalF_{i+1},...,\ccalF_f\} $. 
	Also assume the automaton returns to $ q_D $ but $ \mathds{A}\not =\{\ccalF_{i},\ccalF_{i+1},...,\ccalF_f\} $ then at least one accepting set $ \ccalF_j,~j > i $ has been removed from $ \mathds{A} $ (Note that an accepting set $\ccalF_k,~k<i$ cannot be added since the set of satisfied G-subformulas monotonically expands). This essentially means that $ \square\zeta_j[\mathcal{G}]$ is already checked while $\square\zeta_i[\mathcal{G}]$ is not checked yet, making $ \square\zeta_j $ a subformula of $ \square\zeta_i $. This violates the ordering of $ \mathcal{G} $ and hence the assumption of $\mathds{A}$ being time-variant is not correct.
	
	\subsection{Proof of Theorem \ref{thm1}}
	We prove this result by contradiction. Consider any policy $\overline{\boldsymbol\mu}$ whose traces satisfy $\phi$ with positive probability. Policy $\overline{\boldsymbol\mu}$ induces a Markov chain $\mathfrak{P}^{\overline{\boldsymbol\mu}}$ when it is applied over the MDP $\mathfrak{P}$. This Markov chain comprises a disjoint union between a set of transient states $\mathcal{T}_{\overline{\boldsymbol\mu}}$ and $h$ sets of irreducible recurrent classes $\mathcal{R}^i_{\overline{\boldsymbol\mu}},~i=1,...,h$ \cite{stochastic}, namely: $	\mathfrak{P}^{\overline{\boldsymbol\mu}}=\mathcal{T}_{\overline{\boldsymbol\mu}} \sqcup \mathcal{R}^1_{\overline{\boldsymbol\mu}} \sqcup ... \sqcup \mathcal{R}^h_{\overline{\boldsymbol\mu}}$. 
	%
	%
	From the accepting condition of the LDBA, traces of policy $\overline{\boldsymbol\mu}$ satisfy $\phi$ with positive probability if and only if $$\exists \mathcal{R}^a_{\overline{\boldsymbol\mu}} ~\mbox{s.t.}~ \forall j\in\{1,...,f\},~\mathcal{F}_j^\mathfrak{P} \cap \mathcal{R}^a_{\overline{\boldsymbol\mu}} \neq \emptyset.$$ The recurrent class $\mathcal{R}^a_{\overline{\boldsymbol\mu}}$ is called an accepting recurrent class. Note that if all $h$ recurrent classes are accepting then traces of policy $\overline{\boldsymbol\mu}$ satisfy $\phi$ with probability one. By construction of the reward function \eqref{def:prodRewMDPi} the agent receives a positive reward $ r $ ever after it has reached an accepting recurrent class as it keeps visiting all the accepting sets $\mathcal{F}_j$ infinitely often. 
	
	There are two other possibilities concerning the remaining recurrent classes that are not accepting. A non-accepting recurrent class, name it $ \mathcal{R}^n_{\overline{\boldsymbol\mu}} $, either (i) has no intersection with any accepting set $ \mathcal{F}_j^\mathfrak{P} $, or (ii) or has intersection with some of the accepting sets but not all of them. In case (i), the agent does not visit any accepting set in the recurrent class and the likelihood of visiting accepting sets within the transient states $ \mathcal{T}_{\overline{\boldsymbol\mu}} $ is zero since $ \mathcal{Q}_D $ is invariant. %
	In case (ii), the agent is able to visit some accepting sets but not all of them. This means that there exist always at least one accepting set $\mathcal{F}_j^\mathfrak{P}$ that has no intersection with $ \mathcal{R}^n_{\overline{\boldsymbol\mu}} $ and after a finite number of times, no positive reward can be obtained, and the re-initialization of $ \mathds{A} $ in Definition \ref{frontier} will never happen. By \eqref{eq:utility}, in both cases, for any arbitrary $ r>0 $, there always exists a $ \gamma $ such that the expected reward of a trace reaching an accepting recurrent class such as $\mathcal{R}^a_{\overline{\boldsymbol\mu}}$ with infinite number of positive rewards, is higher than the expected reward of any other trace. 
	
	Next, assume that the traces of optimal policy $\boldsymbol\mu^*$, defined in~\eqref{eq:policy0}, do not satisfy the property $\phi$. In other words, $\forall \mathcal{R}^i_{{\boldsymbol\mu}^*},~\exists j \in \{1,...,f\},~ \mathcal{F}_j^{\mathfrak{P}} \cap \mathcal{R}^i_{{\boldsymbol\mu}^*} = \emptyset$ and all of the recurrent classes are non-accepting. As discussed in cases~(i) and~(ii) above, the accepting policy $ \overline{\boldsymbol\mu} $ has a higher expected reward than the optimal policy $ {\boldsymbol\mu}^* $ due to expected infinite number of positive rewards in policy $ \overline{\boldsymbol\mu} $. However, this contradicts the optimality of $ {\boldsymbol\mu}^* $ in~\eqref{eq:policy0}, completing the proof.
	
	\subsection{Proof of Theorem \ref{thm2}}
	
	We first review how the satisfaction probability is calculated traditionally when the MDP is fully known and then we show that the proposed algorithm convergence is the same. Normally when the MDP graph and transition probabilities are known, the probability of property satisfaction is often calculated via DP-based methods such as standard value iteration over the product MDP $\mathfrak{P}$~\cite{baier2008principles}. This allows to convert the satisfaction problem into a reachability problem. The goal in this reachability problem is to find the maximum (or minimum) probability of reaching \textit{AMECs}. 
	
	The value function $ V:\mathcal{S}\rightarrow [0,1] $ in value iteration is then initialized to $ 0 $ for non-accepting maximum end components and to $ 1 $ for the rest of the MDP. Once value iteration converges then at any given state $ s\in\mathcal{S} $ the optimal policy $ \boldsymbol\mu^*:\mathcal{S}\rightarrow\mathcal{A}_\mathfrak{P} $ is produced by $\boldsymbol\mu^*({s})=\argmax\limits_{a}\sum\limits_{{s}'\in\mathcal{S}} P({s},a,{s}')V^*({s}'),$
	%
	where $ V^*$ is the converged value function, representing the maximum probability of satisfying the property at state $ s $, i.e. $U^{\boldsymbol\mu^*}(s)$ in our setup. 
	
	The key to compare standard model-checking methods to our method is reduction of value iteration to \textit{basic form}. More specifically, quantitative model-checking over an MDP with a reachability predicate can be converted to a model-checking problem with an equivalent reward predicate which is called the basic form. This reduction is done by adding a one-off (or sometimes called terminal) reward of $ 1 $ upon reaching AMECs \cite{pareto}. Once this reduction is done, Bellman operation is applied over the value function (which represents the satisfaction probability) and policy $ \boldsymbol\mu^* $ maximizes the probability of satisfying the property. 
	
	In the proposed method, when an AMEC is reached, all of the automaton accepting sets have surely been visited by policy $ \boldsymbol\mu^* $ and an infinite number of positive rewards $ r>0 $ will be given to the agent as shown in Theorem \ref{thm1}. 
	
	There are two natural ways to define the total discounted rewards \cite{dis2undis}: (i) to interpret discounting as the coefficient in front of the reward; and (ii) to define the total discounted rewards as a terminal reward after which no reward is given and treat the update rule as if it is undiscounted.
	%
	%
	It is well-known that the expected total discounted rewards corresponding to these methods are the same; see, e.g., \cite{dis2undis}. Therefore, without loss of generality, given any discount factor $ \gamma $, and any positive reward component $ r $, the expected discounted reward for the discounted case (the proposed algorithm) is $ c $ times the undiscounted case (value iteration) where $ c $ is a positive constant. This concludes that maximizing one is equivalent to maximizing the other.
	
	\subsection{Proof of Corollary~\ref{cor2}}
	
	Assume that there exists no policy in $\mathfrak{M}$ whose traces can satisfy the property $\phi$. Construct the induced Markov chain $\mathfrak{P}^{\boldsymbol\mu}$ for any arbitrary policy ${\boldsymbol\mu}$ and its associated set of transient states $\mathcal{T}_{{\boldsymbol\mu}}$ and its $h$ sets of irreducible recurrent classes $\mathcal{R}^i_{{\boldsymbol\mu}}$: $\mathfrak{P}^{\boldsymbol\mu}=\mathcal{T}_{{\boldsymbol\mu}} \sqcup \mathcal{R}^1_{{\boldsymbol\mu}} \sqcup \ldots \sqcup \mathcal{R}^h_{{\boldsymbol\mu}}$.
	%
	%
	By assumption, policy $\boldsymbol\mu$ cannot satisfy the property and thus $\forall \mathcal{R}^i_{\boldsymbol\mu},~\exists j \in \{1,...,f\},~ \mathcal{F}_j^\mathfrak{P} \cap \mathcal{R}^i_{\boldsymbol\mu} = \emptyset$. Following the same logic as in the proof of Theorem~\ref{thm1}, after a limited number of times no positive reward is given to the agent. However, by the convergence guarantees of QL, Algorithm~\ref{alg:stationary} will generate a policy with the highest expected accumulated reward. By construction of the reward function in~\eqref{def:prodRewMDPi}, this policy has the highest number of intersections with accepting sets.
	
	%
	%
	
	\subsection{Counter-example}
	
	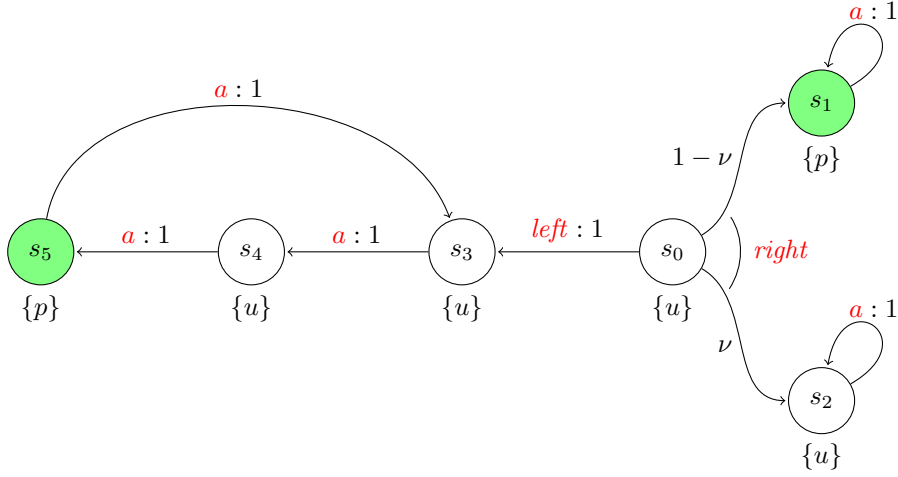
\begin{figure*}[!t]\centering
		\scalebox{1}{
			\begin{tikzpicture}[shorten >=1pt,node distance=2.8cm,on grid,auto] 
			\node[state,label=below:$\{u\}$] (s_0) {$s_0$};
			\node[state,label=below:$\{p\}$,fill=green!50] (s_1) [above right=of s_0]{$s_1$};
			\node[state,label=below:$\{u\}$] (s_2) [below right=of s_0]{$s_2$};
			\node[state,label=below:$\{u\}$] (s_3) [left=of s_0]{$s_3$};
			\node[state,label=below:$\{u\}$] (s_4) [left=of s_3]{$s_4$};
			\node[state,label=below:$\{p\}$,fill=green!50] (s_5) [left=of s_4]{$s_5$};
			
			\draw[->] (s_0) [out=30,in=180] to coordinate [pos=0.1] (r_1) node [left] {$1-\nu$} (s_1);
			\draw[->] (s_0) [out=-30,in=180] to coordinate [pos=0.1] (r_2) node [left] {$\nu$} (s_2);
			
			\draw[->] (s_1) [out=30,in=80,loop] to coordinate[pos=0.1](aa_1) node [above] {$\red{a}:1$} (s_1);
			\draw[->] (s_2) [out=30,in=80,loop] to coordinate[pos=0.1](aa_2) node [above] {$\red{a}:1$} (s_2);
			
			\draw[->] (s_0) -- node [above] {$\red{\mathit{left}}:1$} (s_3);
			\draw[->] (s_3) -- node [above] {$\red{a}:1$} (s_4);
			\draw[->] (s_4) -- node [above] {$\red{a}:1$} (s_5);
			
			\draw[->] (s_5) [out=80,in=110] to coordinate[pos=0.1](bb) node [above] {$\red{a}:1$}(s_3);
			
			\path pic[draw, angle radius=9mm,"$\red{\mathit{right}}$",angle eccentricity=1.6] {angle = r_2--s_0--r_1};
			
			\end{tikzpicture}}
		\caption{Example Product MDP with $\mathcal{AP}=\{p,u\}$ with $\phi=\square\lozenge p$}\label{mdp}
	\end{figure*}
	
	We would like to emphasise that in this work and \cite{hasanbeig2018logically} $0\leq\gamma\leq 1$ due to the fact the algorithm that we proposed is ``episodic'' and thus, covers the un-discounted case as well. This has been unfortunately overlooked in \cite{hahn}. In the following we examine the general cases of discounted and un-discounted learning and we show that our algorithm, which is episodic, is able to output the correct action for the example provided in \cite{hahn} (Fig. \ref{mdp}). For the sake of generality, we have parameterised the probabilities associated with action $\mathit{right}$ and $\mathit{left}$ with $1-\nu$ and $\nu$, respectively.
	
	Recall that for a policy $\boldsymbol\mu:\mathcal{S}\rightarrow\mathcal{A}$ on an MDP $\mathfrak{M}$, and given a reward function $R$, the expected discounted reward at state $s$ by taking action $a$ is defined as \cite{sutton}:
	
	\begin{equation}\label{expected_return}
	{U}^{\boldsymbol\mu}(s,a)=\mathds{E}^{\boldsymbol\mu} [\sum\limits_{n=0}^{\infty} \gamma^n~ R(s_n,\boldsymbol\mu(s_n))|s_0=s, a_0=a],
	\end{equation}
	
	where $\mathds{E}^{\boldsymbol\mu}[\cdot]$ denotes the expected value by following policy ${\boldsymbol\mu}$, and $s_1,a_1,...,s_n,a_n$ is the sequence of state-action pairs generated by policy $\mathit{Pol}$ up to time step $n$.  
	
	We would like to show that for some $\gamma \in [0,1]$, ${U}^{{\boldsymbol\mu}}(s_0,\mathit{left})>{U}^{{\boldsymbol\mu}}(s_0,\mathit{right})$. From \eqref{expected_return}, at state $s_0$, the expected return for each action is:
	\begin{equation}\label{total}
	\begin{aligned}
	&{U}^{{\boldsymbol\mu}}(s_0,\mathit{right})=(1-\nu)[r+\gamma r + \gamma^2 r + ...]\\
	&{U}^{{\boldsymbol\mu}}(s_0,\mathit{left})=\gamma^2 r + \gamma^5 r + \gamma^8 r + ...
	\end{aligned}
	\end{equation}
	Notice that ${\boldsymbol\mu}$ has no effect on the expected return after the agent chose to go right or left as there is only one action available in subsequent states. Let us first consider ${U}^{{\boldsymbol\mu}}(s_0,\mathit{right})$. The RHS is a geometric series with the initial term $(1-\nu)r$ and ratio of $\gamma$. Thus,
	\begin{equation}\label{right}
	\begin{aligned} {U}^{{\boldsymbol\mu}}(s_0,\mathit{right})=(1-\nu)r \dfrac{1-\gamma^n}{1-\gamma}.
	\end{aligned}
	\end{equation}
	
	The expected return ${U}^{{\boldsymbol\mu}}(s_0,\mathit{left})$ is also a geometric series such that:
	
	\begin{equation}\label{left}
	\begin{aligned} {U}^{{\boldsymbol\mu}}(s_0,\mathit{left})=\gamma^2 r \dfrac{1-\gamma^{3n}}{1-\gamma^3}.
	\end{aligned}
	\end{equation}
	
	Consider two cases (1) $0\leq\gamma<1$, and (2) $\gamma=1$.
	
	In the first case $0\leq\gamma<1$, as $n\rightarrow\infty$, $\gamma^n\rightarrow0$ and $\gamma^{3n}\rightarrow0$ and therefore, the following inequality can be solved for $\gamma$:
	
	\begin{equation}\label{ineq}
	\begin{aligned}  
	&\dfrac{\gamma^2 r}{1-\gamma^3}>\dfrac{(1-\nu)r}{1-\gamma}~\longrightarrow ~\dfrac{\gamma^2}{1+\gamma+\gamma^2}>1-\nu~\longrightarrow~\\
	~\\
	&\left\{
	\begin{array}{lr}
	\gamma<\dfrac{(-\sqrt{1/\nu^2 + 2/\nu - 3} - 1) \nu + 1}{2\nu},\\
	~\\
	~or\\
	~\\
	\gamma>\dfrac{(\sqrt{1/\nu^2 + 2/\nu - 3} - 1) \nu + 1}{2\nu}.
	\end{array}
	\right.
	\end{aligned}
	\end{equation}
	Thus, for some $\nu\in[0,1]$, the discounted case $0\leq\gamma<1$ is sufficient if $\gamma>{(\sqrt{1/\nu^2 + 2/\nu - 3} - 1) \nu + 1}/{2\nu}$. However, it is possible that for some $\nu\in[0,1]$ both conditions push $\gamma$ to be outside of its range of $0\leq\gamma<1$ in the first case. Therefore, in the learning algorithm $\gamma$ needs to be equal to $1$, which brings us to the second case, that is allowed in our work thanks to the episodic nature of our algorithm. 
	
	Note that when $\gamma=1$ we cannot derive \eqref{ineq} since $\lim_{n\rightarrow\infty} \gamma^n=\lim_{n\rightarrow\infty} \gamma^{3n}=1$, and also $1-\gamma=0$ cannot be cancelled from both sides of the inequality. Further to this, \eqref{right} and \eqref{left} become undefined when $\gamma=1$. From \eqref{total} though, we know with $\gamma=1$, the summations go to infinity as $n\rightarrow\infty$. The question is, can we show that ${U}^{{\boldsymbol\mu}}(s_0,\mathit{left})>{U}^{{\boldsymbol\mu}}(s_0,\mathit{right})$. 
	
	Recall that the convergence of QL is asymptotic and if we can show that ${U}^{{\boldsymbol\mu}}(s_0,\mathit{left})>{U}^{{\boldsymbol\mu}}(s_0,\mathit{right})$ after a number of episodes, then essentially our algorithm can output the correct result and will choose action $\mathit{left}$ once QL has converged. To prove this claim let us consider the following limit as we push $\gamma$ towards $1$:
	
	\begin{align}
	\begin{aligned} &\lim\limits_{\gamma\rightarrow1} \dfrac{{U}^{{\boldsymbol\mu}}(s_0,\mathit{left})}{{U}^{{\boldsymbol\mu}}(s_0,\mathit{right)}}=\lim\limits_{\gamma\rightarrow1}\dfrac{\gamma^2 r \dfrac{1-\gamma^{3n}}{1-\gamma^3}}{(1-\nu)r \dfrac{1-\gamma^n}{1-\gamma}}=\\
	&\lim\limits_{\gamma\rightarrow1}\dfrac{\gamma^2 \bcancel{r} \dfrac{\cancel{(1-\gamma^{n})}(1+\gamma^n+\gamma^{2n})}{\cancel{(1-\gamma)}(1+\gamma+\gamma^2)}}{(1-\nu)\bcancel{r} \dfrac{\cancel{1-\gamma^n}}{\cancel{1-\gamma}}}=\dfrac{\gamma^2}{1-\nu}
	\end{aligned}
	\end{align}
	In case when $\nu=0,~1-\nu=1$ then the limit is $1$, namely the algorithm is indifferent between choosing $\mathit{left}$ or $\mathit{right}$. This matches with the MDP as well since going to either direction does not change the optimality of the action when $1-\nu=1$. However, if $0<\nu\leq1$ then the limit is always greater than one, meaning that the expected return for taking left is greater than taking right after some finite number of episodes. 
	
\end{document}